\documentclass[fleqn,usenatbib]{mnras}
\usepackage[T1]{fontenc}
\usepackage{ae,aecompl}
\usepackage{alltt}

\usepackage{graphicx}	% Including figure files
\usepackage{amsmath}	% Advanced maths commands
\usepackage{amssymb}	% Extra maths symbols

\newcommand{\RN}[1]{\textup{\uppercase\expandafter{\romannumeral#1}}}

\newcommand{\Msun}{\rm M_{\odot}}

\title[Orbital Properties of Dark Matter Haloes]{Baryonic impact on the dark matter orbital properties of Milky Way-sized haloes}

\author[Q. Zhu et al.]
{Qirong Zhu$^{1,2}$,\thanks{E-mail: qxz125@psu.edu}
Lars Hernquist$^{3}$,
Federico Marinacci$^{4}$, 
Volker Springel$^{5,6}$, and
\newauthor  
Yuexing Li$^{1,2}$ \vspace{0.5cm}\\
\parbox{\textwidth}
{\small $^{1}$Department of Astronomy \& Astrophysics, The Pennsylvania State University, 525 Davey Lab, Penn State, PA, 16802, USA\\
$^{2}$Institute for Cosmology and Gravity, The Pennsylvania State University, University Park, PA, 16802, USA\\
$^{3}$Harvard-Smithsonian Center for Astrophysics, Harvard University, 60 Garden Street, Cambridge, MA 02138, USA\\
$^{4}$Department of Physics, Kavli Institute for Astrophysics and Space Research, MIT, Cambridge, MA 02139, USA \\
$^{5}$Heidelberg Institute for Theoretical Studies, Schloss-Wolfsbrunnenweg 35, 69118 Heidelberg, Germany\\
$^{6}$Zentrum f\"{u}r Astronomie der Universit\"{a}t Heidelberg, ARI, M\"{o}nchhofstr. 12-14, 69120 Heidelberg, Germany}}

\date{Accepted XXX. Received YYY; in original form ZZZ}

\pubyear{2016}

\begin{document}
\label{firstpage}
\pagerange{\pageref{firstpage}--\pageref{lastpage}}
\maketitle

% Abstract of the paper
\begin{abstract}
We study the orbital properties of dark matter haloes by combining a
spectral method and cosmological simulations of Milky Way-sized
galaxies. We compare the dynamics and orbits of individual dark matter
particles from both hydrodynamic and $N$-body simulations, and find
that the fraction of {box, tube} and {resonant} orbits of the dark matter halo
decreases significantly due to the effects of baryons. In
particular, the central region of the dark matter halo in the
hydrodynamic simulation is dominated by regular, short-axis tube
orbits, in contrast to the chaotic, box and thin orbits dominant in
the $N$-body run. This leads to a more spherical dark matter halo in
the hydrodynamic run compared to a prolate one as commonly seen in the
$N$-body simulations. Furthermore, by using a kernel based density
estimator, we compare the coarse-grained phase-space densities of dark
matter haloes in both simulations and find that it is lower by
$\sim0.5$ dex in the hydrodynamic run due to changes in the angular
momentum distribution, which indicates that the baryonic process that
affects the dark matter is irreversible.  Our results imply that
baryons play an important role in determining the shape, kinematics
and phase-space density of dark matter haloes in galaxies.
\end{abstract}

% Select between one and six entries from the list of approved keywords.
% Don't make up new ones.


%%%%%%%%%%%%%%%%%%%%%%%%%%%%%%%%%%%
%%%%%%%%%%%%%%%%% BODY OF PAPER %%%%%%%%%%

%%%%%%%%%%%%%%%%% INTRODUCTION  %%%%%%%%%%
\section{Introduction}

Within the cold dark matter (CDM) cosmology framework, it has been
long known that collapsed haloes show a cuspy density distribution called 
Navarro-Frenk-White (NFW) profile \citep[][]{NFW}, as well as strongly triaxial shapes
\citep[][]{Jing2002}. For a self-gravitating system, there is no
equilibrium in a thermodynamic sense \citep{LyndenBell1967} that the
system can rest on \citep{BinneyTremaine}. The present-day properties
of DM haloes depend on their evolutionary paths starting from cosmological
initial conditions (see the example given by \citealt{Arad2005}).  In the
hierarchical growth picture, triaxial haloes naturally arise from
halo mergers with the remnant shape closely connected to the orbits of
the progenitors. \cite{Moore2004} found that mergers on radial orbits
produce prolate systems while circular orbits result in oblate
ones. However, the steep density slope in the centre as in the NFW
profile \citep{NFW} does not necessarily originate from mergers. It was
suggested from 1-D studies that a single mode gravitational collapse
naturally produces a power-law density profile in the centre
\citep[e.g.,][]{Binney2004, Schulz2013, Colombi2014}, although the
central density slopes derived from these studies differ from each
other. Steep cusps, on the other hand, are persistent during mergers
\citep{Dehnen2005} which possibly leads to an attractor
\citep{Syer1998, Loeb2003}, as suggested by numerical simulations
\citep{Angulo2016}.

While recent works by \cite{Vogelsberger2008} and \cite{Abel2012} made
progress in the study of DM using fine-grained phase-space density,
our knowledge of the halo properties are mostly based upon
collisionless $N-$body simulations. It has been suggested that baryons
have a profound effect on the distribution of dark matter
\citep[][]{Navarro1996, Lackner2010}. The intrinsic shape of DM haloes is closely
linked with the orbital properties of its constituents
\citep[e.g.][]{Schwarzschild1979, Barnes1996, HolleyBockelmann2001, Jesseit2005,
  Hoffman2010}. In a 3-D non-spherical system such that the potential
is not fully integrable, only energy is conserved. Orbits in such
potentials are usually not closed and several major orbital families
have been long recognized \citep[e.g.][]{Schwarzschild1979,
  Statler1987}. Among these orbit families, box orbits, which are the
backbone of triaxial systems, can pass quite close to the potential
centre and be deflected by a supermassive black hole
\citep{Gerhard1985}.  From this effect, over time, an original
triaxial system can be shaped into a more spherical one. In galaxy
mergers with gas cooling and star formation included, orbital families
are also found to be sensitively dependent on the gas content
\citep{Barnes1996, Hoffman2010}.

Recently, we studied the baryonic effects on DM properties by
comparing a cosmological hydrodynamic simulation (C-4) of a Milky Way-sized
galaxy with its dark matter only (DMO) counterpart \citep{Zhu2016}. We
found the shape of the DM halo in the hydrodynamic simulation appears
to be nearly spherical, in contrast to the oblate shape in the DMO
run. The more spherical DM halo in the hydrodynamic simulation agrees
with other studies using different numerical techniques and models.
Some of the previous studies \citep[e.g.][]{Maccio2007,
  Debattista2008, Valluri2010, Bryan2012} have focused on the orbital
properties of the halo using spectral techniques \citep{Binney1982,
  Laskar1993}. While these studies agreed that a significant fraction
of box orbits are replaced by short axis tube orbits when a baryonic
disk is present, whether or not such a process is
reversible remains under debate. 

Using a controlled experiment including a growing and evaporating disk
potential, \cite{Valluri2010} concluded that the change induced by
baryons on the DM particle orbits critically depends on the radial
distribution of baryons. More centrally concentrated baryons introduce
changes in a more irreversible fashion. Thanks to the rapid 
developments within the past few years \citep[e.g.][]{Guedes2011,
  Agertz2011, Aumer2013, Marinacci2014}, we can now study DM haloes
from cosmological hydrodynamic simulations to better understand the
impact of baryons on the DM distribution.

In this study, we apply an automatic orbit classification code called
{\sc smile}\footnote{The source code of {\sc smile} is 
available at \url{td.lpi.ru/~eugvas/smile/}.} \citep{smile}, 
which uses a spectral method, to analyze the DM
particle orbits of the Milky Way-sized halo studied in
\cite{Zhu2016}. Compared with previous studies, we use a more accurate
representation of the potential field based on our cosmological
simulations and a high order integration scheme which greatly
improves the accuracy of indications of chaos. Moreover, the
classification of orbits in our study further distinguishes resonant
and thin orbits \citep{Merritt1999} from a broad box orbit family. 
Furthermore, the cosmological hydrodynamic simulation makes it
possible to follow the build-up of the baryonic disk in detail, which
itself has a manifestly complex angular momentum evolution
\citep[e.g.][]{Grand2016}.  Possible coupling between the angular
momentum of the baryonic disk and the DM halo is a critical process 
that determines whether or not the impact induced by baryons on the 
DM is reversible. Such a process, however, was not properly modeled 
in previous studies using a rigid body potential  \citep{Debattista2008, Valluri2010}.

This paper is organized as follows. We describe our methods in
Section~\ref{sec:methods} and present the main results in
Section~\ref{sec:results}. We discuss resolution effects and the
triaxial symmetry assumption on the orbital properties in
Section~\ref{sec:discussions}, and summarize our findings in
Section~\ref{sec:conclusions}.

%%%%%%%%%%%%%%%%% METHODS %%%%%%%%%%%%%%%%%%

\begin{figure*}
\begin{center}
\begin{tabular}{ccc}
\resizebox{0.66\columnwidth}{!}{\includegraphics{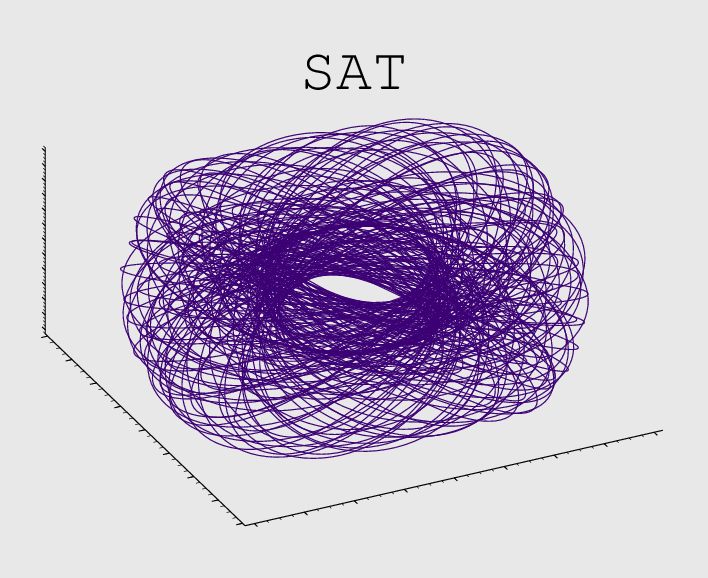}}
\resizebox{0.66\columnwidth}{!}{\includegraphics{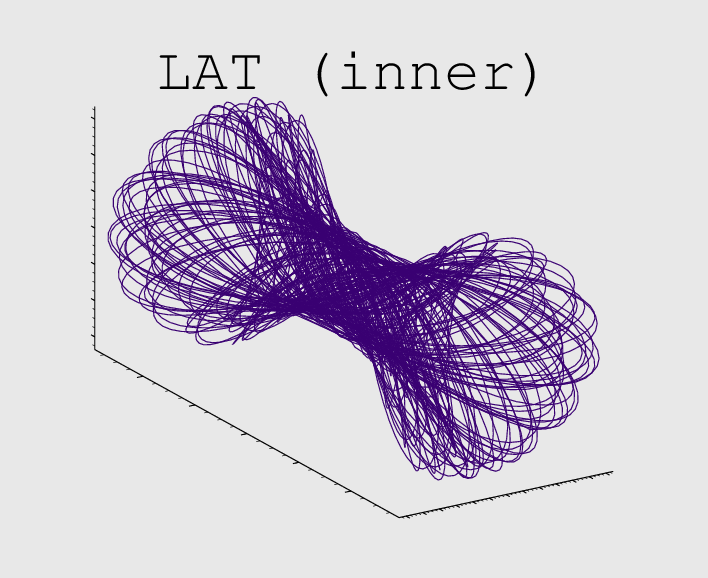}}
\resizebox{0.66\columnwidth}{!}{\includegraphics{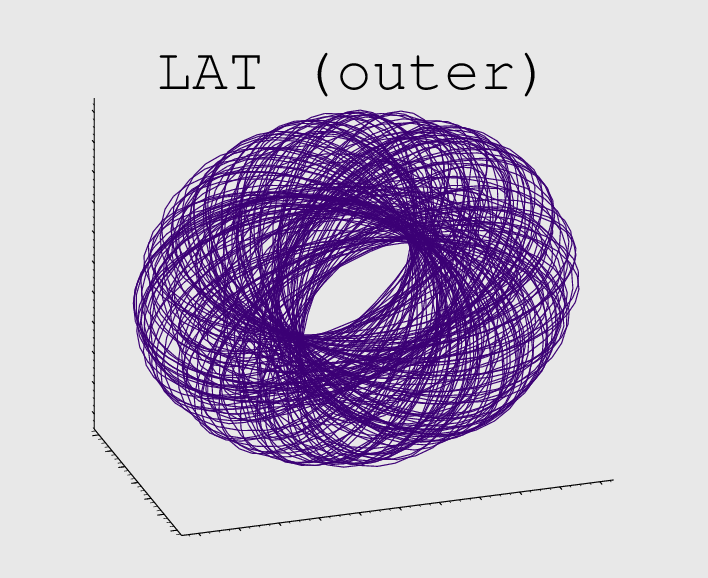}}\\
\resizebox{0.66\columnwidth}{!}{\includegraphics{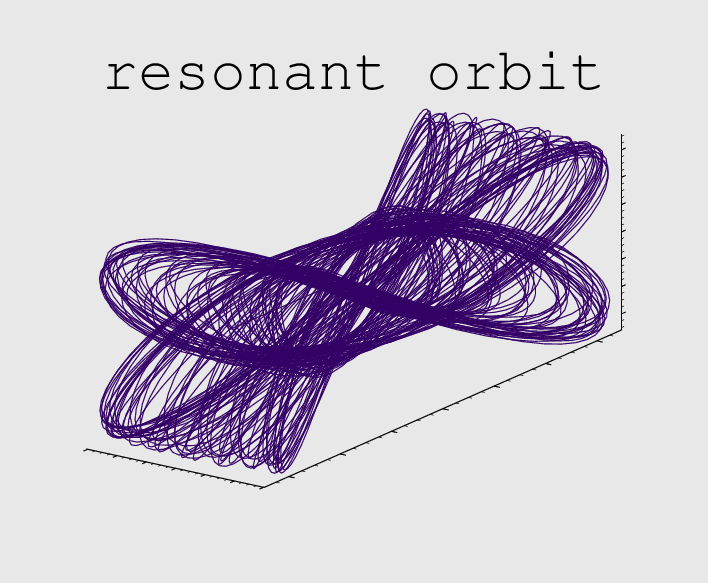}}
\resizebox{0.66\columnwidth}{!}{\includegraphics{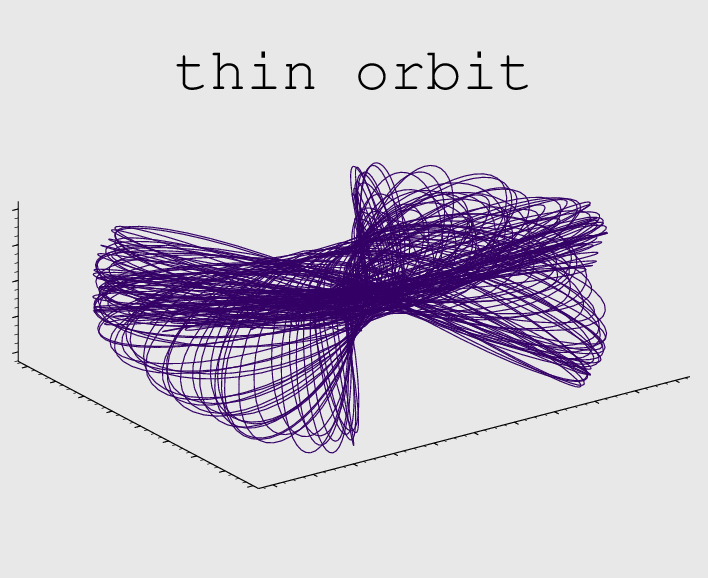}}
\resizebox{0.66\columnwidth}{!}{\includegraphics{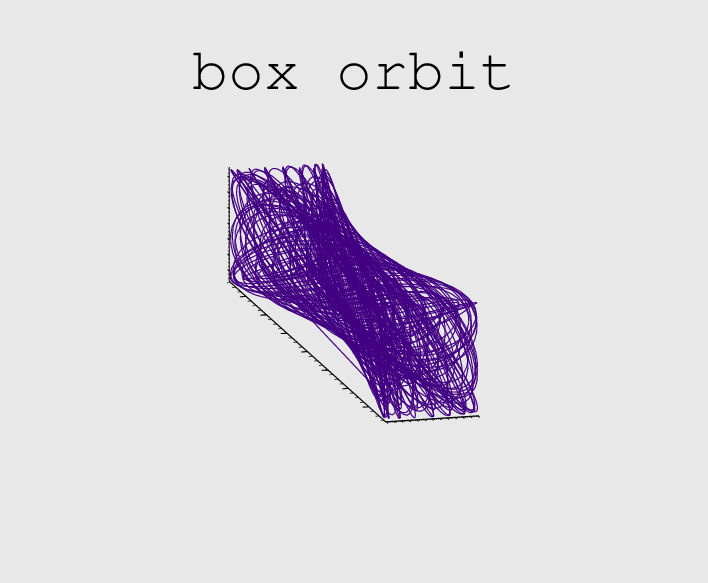}}\\
\end{tabular}
\end{center}
\caption{\label{fig:orbit_family} Orbital families considered in this study. From top left to bottom right: a SAT, an inner LAT, an outer LAT, a (2:$*$:3)-resonant orbit, a thin orbit and a regular box orbit. } 
\end{figure*}

\begin{figure*}
\begin{center}
\begin{tabular}{cc}
\resizebox{0.9\columnwidth}{!}{\includegraphics{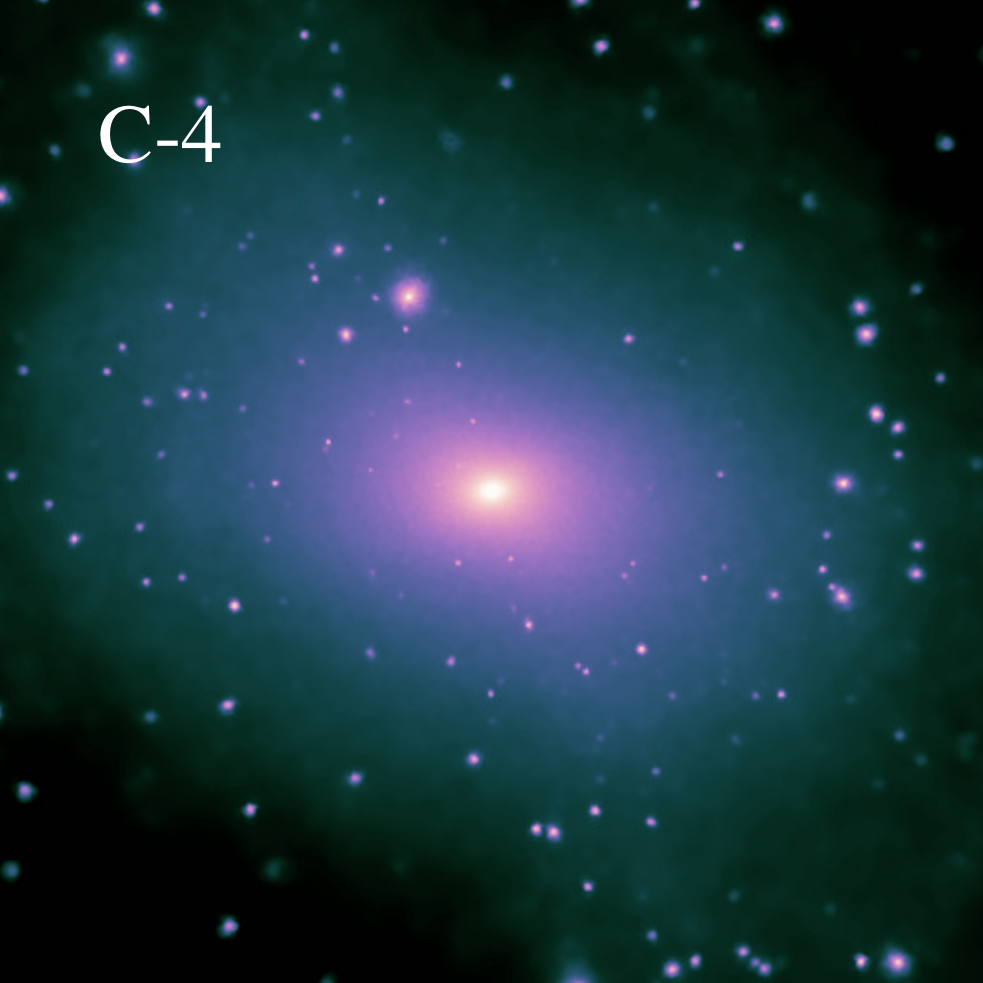}}
\resizebox{0.9\columnwidth}{!}{\includegraphics{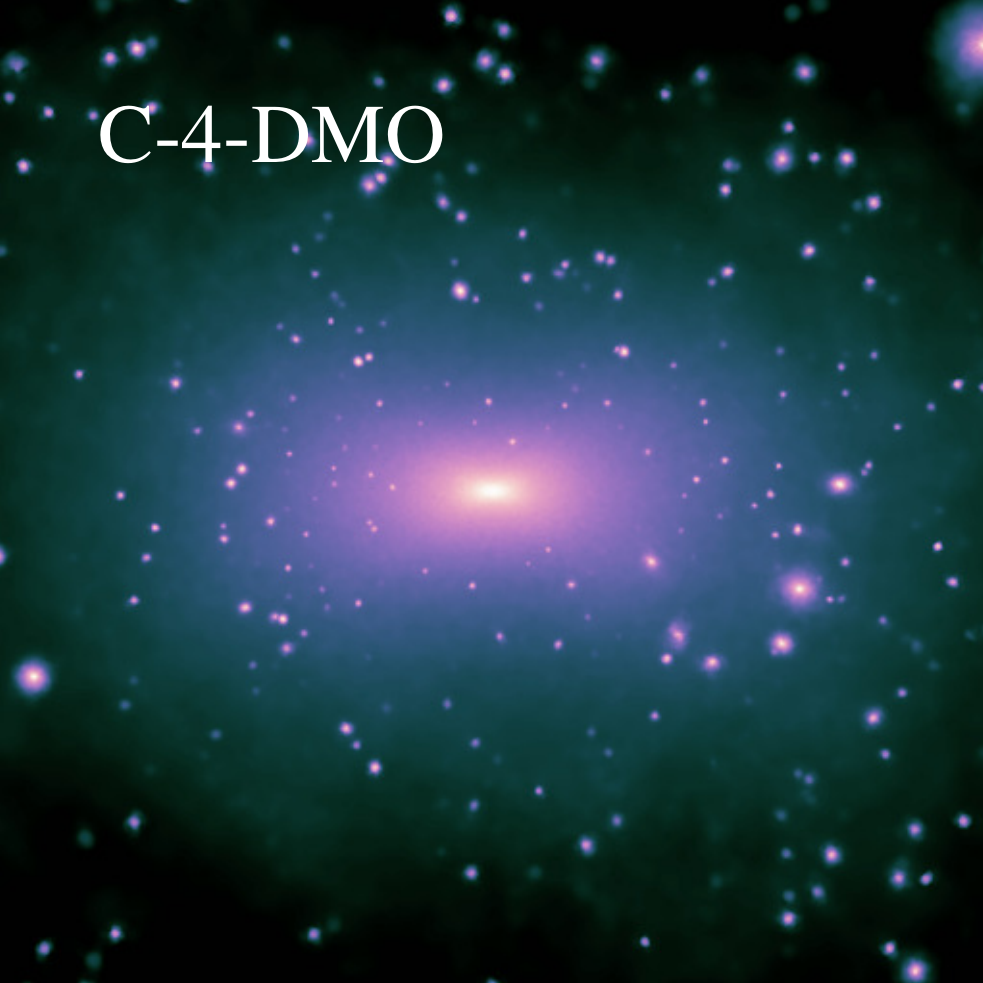}}\\
\resizebox{0.9\columnwidth}{!}{\includegraphics{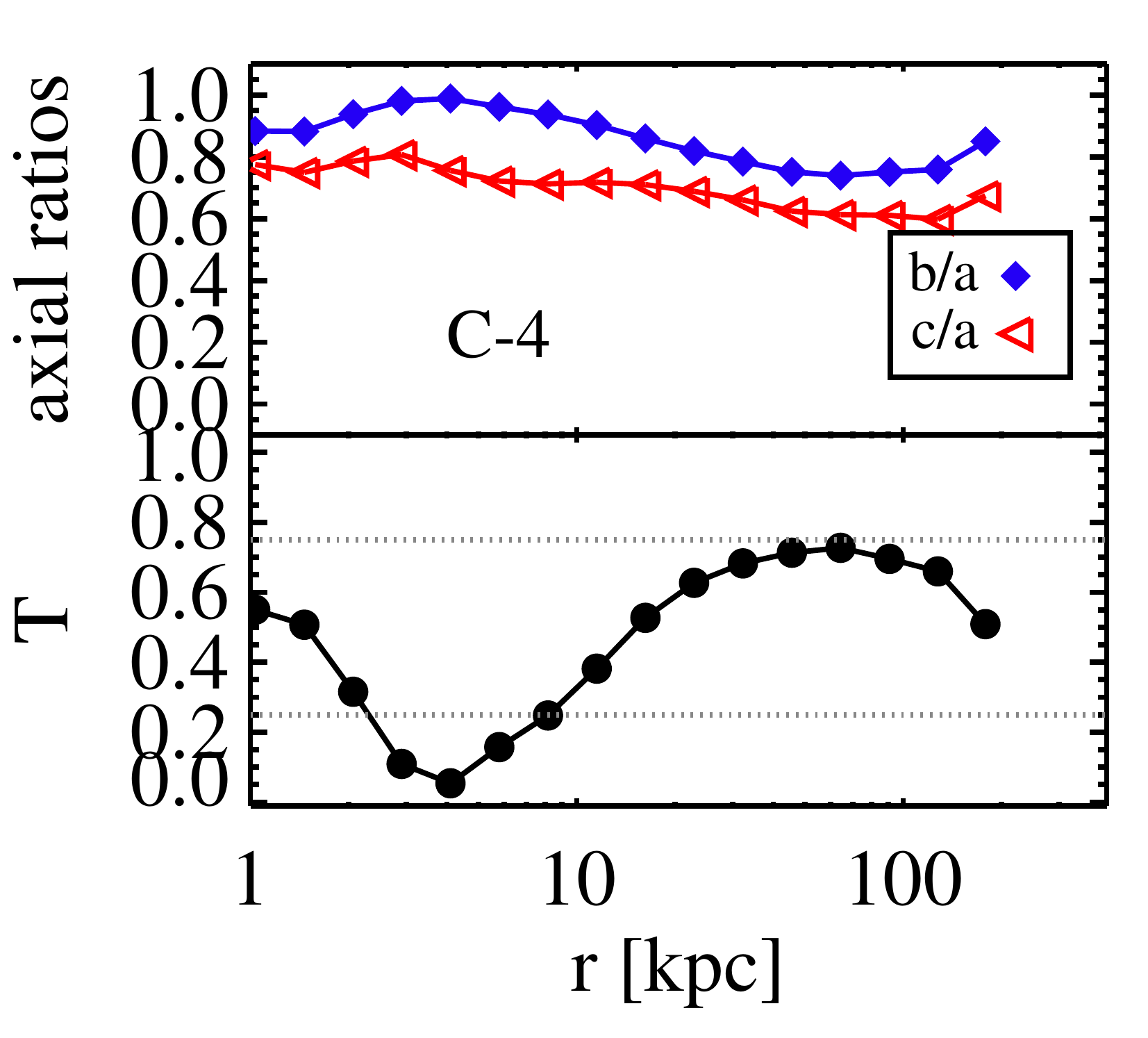}}
\resizebox{0.9\columnwidth}{!}{\includegraphics{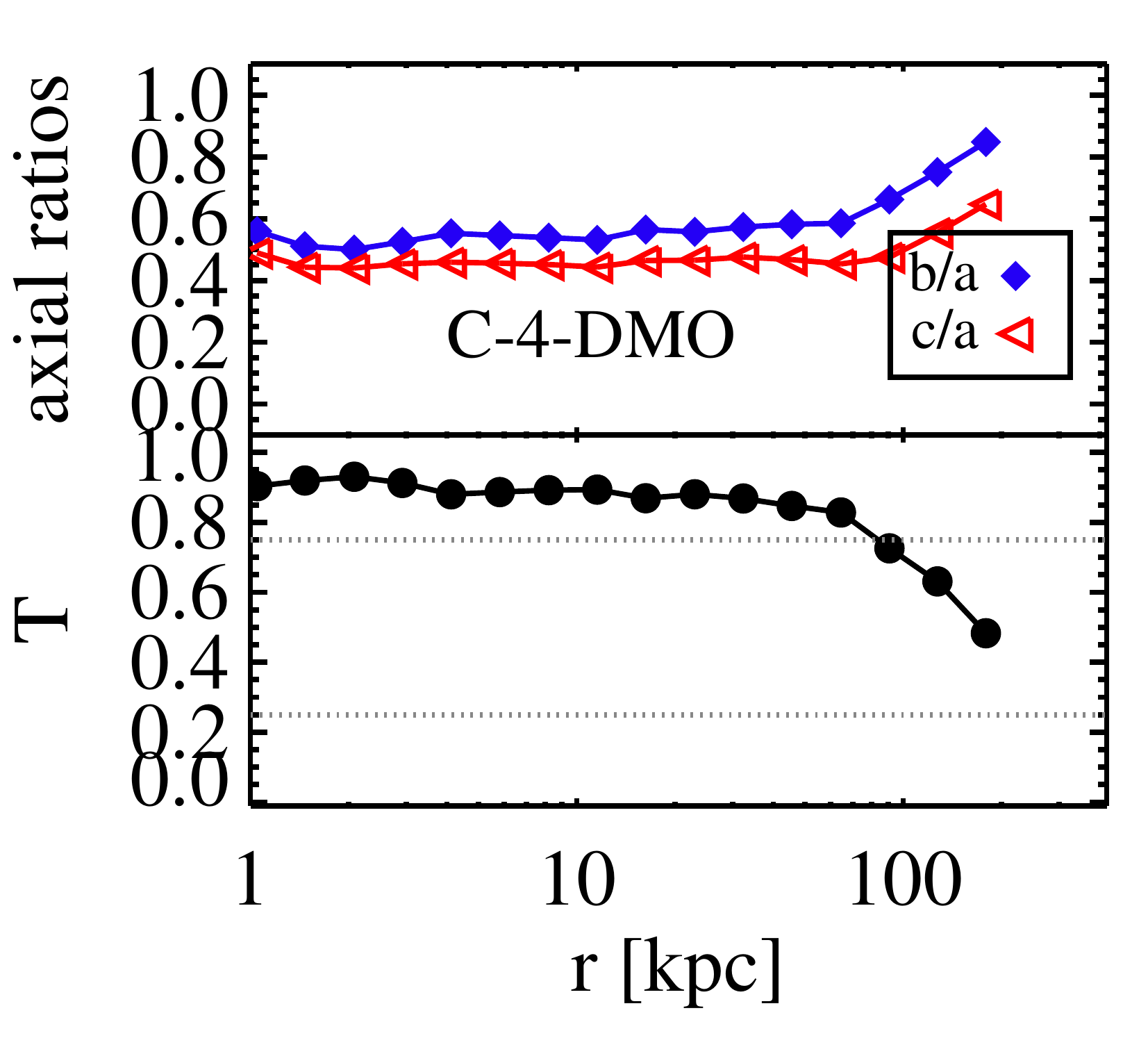}}\\
\end{tabular}
\caption{\label{fig:dmmaps} A comparison of dark matter halo shape between the hydrodynamic and DMO simulations of the Aq-C halo. Top panel: the projected DM density maps of Aq-C haloes in the $x-z$ plane. Bottom panel: the corresponding halo shape parameters in terms of axial ratios ($b/a$ and $c/a$), and triaxiality parameter ($T\equiv(a^2-b^2)/(a^2-c^2)$)  as a function of galactocentric distance $r$. }
\end{center}
\end{figure*}

\begin{figure*}
\begin{center}
\begin{tabular}{cc}
\resizebox{\columnwidth}{!}{\includegraphics{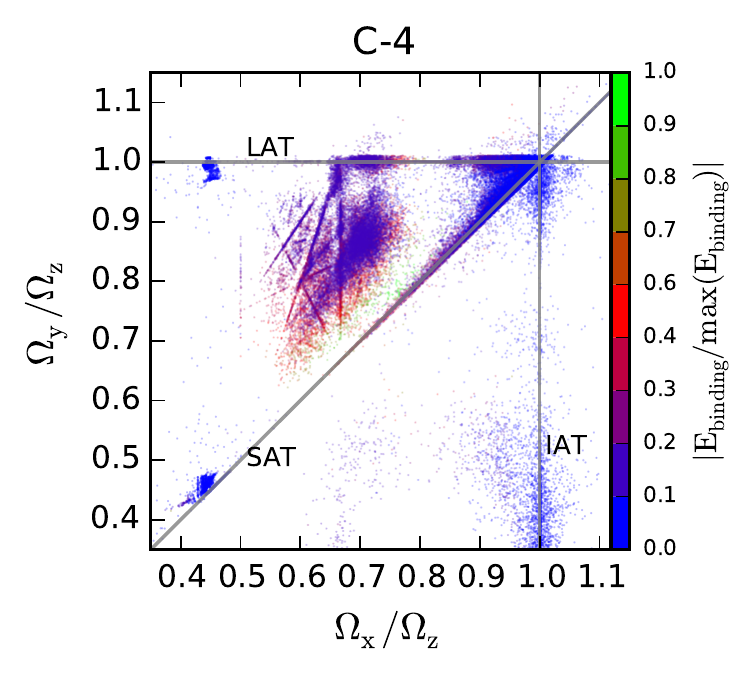}}
\resizebox{\columnwidth}{!}{\includegraphics{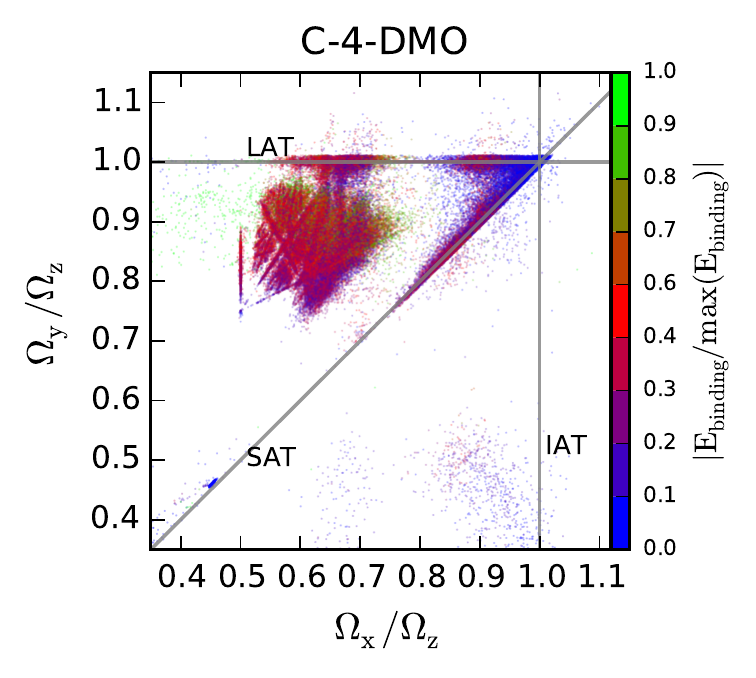}}\\
\resizebox{\columnwidth}{!}{\includegraphics{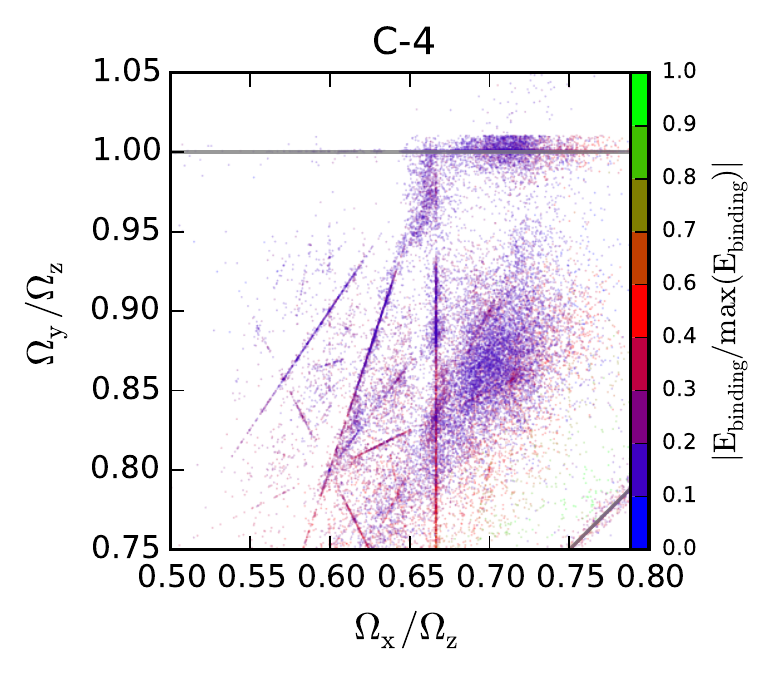}}
\resizebox{\columnwidth}{!}{\includegraphics{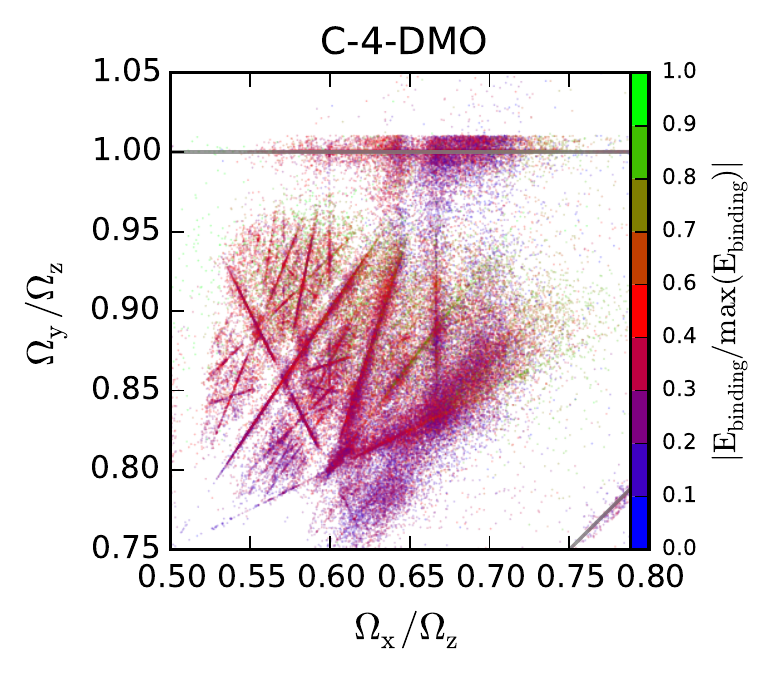}}\\
\end{tabular}
\end{center}
\caption{\label{fig:frequency_map} Frequency maps in the plane of $\Omega_x/\Omega_z$ and $\Omega_x/\Omega_z$ for the Aq-C-4 halo from both hydrodynamic (left column) and DMO (right column) simulations, respectively. The top panels show a larger range than the bottom ones. The dots indicate individual orbits color-coded by their binding energy $E_{\rm binding} = \Phi(r) + \mathbf{v}^2/2$. The three gray solid lines denote the locations for SAT, IAT and LAT orbits, which satisfy $\Omega_x\approx\Omega_y$, $\Omega_x\approx\Omega_z$ and $\Omega_y\approx\Omega_z$, respectively. The lower panels show zoom-in regions of the frequency maps shown in the top panel. For the DMO simulation, numerous thin and resonant orbits are visible in this region. For comparison, the orbits in the hydrodynamic simulation are sparsely populated in the same region.}
\end{figure*}

\section{Methods}
\label{sec:methods}

\subsection{Cosmological simulations}
\label{subsec:sims}

In this study, we use both hydrodynamical and dark matter
only (DMO) cosmological simulations of the Milky Way-sized halo Aq-C-4
from \cite{Zhu2016} (referred to as the C-4 simulation hereafter). Both
simulations have a mass resolution of $3.0 \times 10^{5}\, \Msun$ for
the DM component, while the hydrodynamic one has a mass resolution of $5.0
\times 10^{4}\, \Msun$ for gas and stars, and it includes a
comprehensive list of baryonic processes, such as metal cooling, star
formation, feedback from both supernovae and active galactic nuclei,
and stellar evolution. We have also included a set of simulations with a 
mass resolution 8 times lower than that of the C-4 runs 
(referred to as C-5 hereafter) to study the effects of resolution on the DM
properties.

\subsection{Orbit classification}
\label{subsec:orbitclassfication}

We start from the snapshots at $z = 0$ from the cosmological simulations
to build a smooth potential model that we then use to follow the
orbits of individual DM particles. We first locate the potential
minimum and reorient the entire system to the principle axes
determined by the mass distribution within 20 kpc of the
potential minimum. For the hydrodynamic simulation,
the stellar and gaseous components are included. The new $z$-direction
 in the C-4 simulation aligns almost perfectly with the orientation of its stellar disc. 

We use the spherical-harmonic expansion method in
{\sc smile} using terms up to
$n_{\rm rad} = 12$ in the radial direction and $l_{\rm max} = 6$, which
are suitable for the total particle numbers in the halo.  We further
assume triaxial symmetry which leaves the coefficient of
$Y^{m}_l(\theta, \phi)$ nonzero for even $l$ and $m$ terms. The
above procedure closely follows the self-consistent field (SCF) method
of \cite{Hernquist1992} with the radial functional form replaced by a
non-parametric density estimator using splines, which increases the
accuracy of the approximations to the underlying potential/force field.
In total, 
$\sim8,700,000$ DM particles together with $\sim3,800,000$ gas particles and 
$\sim2,300,000$ star particles are evolved for C-4 simulation.
We then use the position and velocity information of the DM particles
selected from the snapshots as ``tracers". In total, we select all the
$\sim300,000$ DM particles within $500$ kpc from the galaxy center
for C-4 simulation (both hydrodynamic and DMO) and $\sim35,000$ 
particles for C-5 simulation (which is about one eighth of the 
number in C-4 simulation). With the default eighth order 
Runga-Kutta algorithm and adaptive step sizes, we integrate each
particle with 150 full orbits and the fractional error in energy
conservation for the entire duration is below $10^{-5}$. The
conservation of energy is a crucial step for the classification
of different orbit families.

Following \cite{Binney1982}, the coordinates of each orbit 
are analyzed as a function of time using Fourier spectra, 
\begin{equation}\\
x(t)   = \sum_{k} A_{k} \exp(i\omega_k t).
\end{equation}

The {\sc smile} code employs the automatic method proposed by
\cite{Carpintero1998} to identify the location of peaks and extracts
their amplitude in the discrete Fourier transform of the
coordinates \footnote{There are some notable differences between 
the orbit classification in {\sc smile} and the original \cite{Carpintero1998} 
paper. {\sc smile} enforces the following relation 
$$\Omega_x \le \Omega_y \le \Omega_z.$$
As a result, $\Omega_x, \Omega_y, \text{or } \Omega_z$ 
are not necessarily associated with the largest peaks 
(could be the second largest) for some resonant orbits 
while \cite{Carpintero1998} selects the the largest peaks in the spectra. 
In addition, there is no separate `chaotic' orbit family in {\sc smile}.}.  
For regular orbits, at most three frequencies are
linearly independent while the frequency of all the other spectral
lines can be expressed as the combination of the three fundamental
frequencies $\Omega_x$, $\Omega_y$ and $\Omega_z$ as $ \omega_k = l
\Omega_x + m\Omega_y + n \Omega_z$, where $(l, m, n)$ are integer
triplets.

Subsequently, the classification of orbits is based on whether or not
$\Omega_x$, $\Omega_y$ and $\Omega_z$ are commensurable. If 
the frequencies are not related through integer triplets, 
we have box orbits. On the other hand, they are called ($l, m, n$) thin 
orbits if the three fundamental frequencies satisfy the following resonant 
condition: 
\begin{equation}\\
l\Omega_x + m\Omega_y + n \Omega_z = 0, 
\label{eq:triple_resonant_relation}
\end{equation}
as they are confined on a membrane in configuration space \citep{Merritt1999}.

Among thin orbits we have ``resonant orbits" where either $l$, $m$ or
$n$ is 0. The most common subclass of resonant orbits is known as
`tube orbits' which correspond to a 1:1 resonance. We refer to the
orbits with $\Omega_x\approx\Omega_y$ as short axis tubes (SAT, or
Z-tubes) and those with $\Omega_y\approx\Omega_z$ as long axis tubes
(LAT, or X-tubes). SAT and LAT orbits conserve the sign of the $z$ and
$x$ components of angular momentum and hence they contribute net
rotation around the $z$ and $x$ axes. Loop orbits around the
intermediate axis (IAT) in perfect ellipsoidal potentials are
unstable. Long axis tubes are further classified as inner and outer
long-axis tubes \citep[][]{Statler1987, BinneyTremaine}. In this study,
we consider the following most common orbital families: (1) SAT, (2)
inner LAT, (3) outer LAT, (4) resonant orbits, (5) thin orbits, and
(6) boxes, as illustrated in Figure~\ref{fig:orbit_family}.

If the spectral lines cannot be expressed as the linear combination
of at most three fundamental frequencies, they should be labelled as
chaotic according to \cite{Carpintero1998}. 
Since {\sc smile} doest not have a separate `chaotic' orbital 
family, these orbits are often classified under the `box' category and
sometimes can be under `resonant' or `thin' with additional `chaotic' attribute. 
The latter happens if the spectra lines are under the tolerance parameter 
$\epsilon$ for detecting the resonant condition:
\begin{equation}\\
|n_b * \Omega_a  - n_a * \Omega_b| \le \epsilon, 
\label{eq:binary_resonant}
\end{equation}
where $\max{(n_a, n_b)} \le 10$. 
To further study the chaotic properties of the orbits, 
{\sc smile} computes the frequency diffusion
rate $\Delta \Omega$ as the change of frequencies between the first
and second halves of the interval ($\Omega_{k, 1}, \Omega_{k, 2}$):

\begin{equation}\\
\Delta \Omega = \frac{1}{3}\sum_k \frac{|\Omega_{k, 1} - \Omega_{k, 2}|}{(\Omega_{k, 1} + \Omega_{k, 2})/2}.
\label{eq:frequency_diffusion_rate} 
\end{equation}

In addition, the Lyapunov exponent $\Lambda$ using a nearby orbit is
employed to provide a more secure measurement of chaos. All of the
above procedures, potential reconstruction, orbital integration and
classification, are included in the {\sc smile} code where the
classification is automatic. 

The following pseudocode describes
the overall procedure of orbit classification:
\begin{enumerate}
\item Perform FFT of the spatial coordinates and identify spectral lines;
\item Find the leading frequencies: $\Omega_x$,   $\Omega_y$ and  $\Omega_z$ in each coordiantes;
\item Test the binary commensurable relation in Equation~(\ref{eq:binary_resonant}) between each pair of the leading frequencies; 
\begin{enumerate}
\item If no resonant relation in any pairs of the leading frequencies is detected, {\sc smile} 
tests Equation~(\ref{eq:triple_resonant_relation}) to determine whether or not the orbit is a `box' or a `thin' orbit;
\item If there are resonant relations detected in the  pairs of the leading frequencies, the orbit is labeled as `resonant'. 
\end{enumerate}
\item Test whether $L_x$, $L_y$, and $L_z$ switch from positive(negative) values to opposite sign during the course of the integration to determine the orbit is a `tube' orbit or not. 
\item Finally, {\sc smile} tests the relations between the leading frequencies and the rest of the spectral lines in each coordinates to determine whether or not the additional `chaotic' attribute should be added to the orbit classification. 
\end{enumerate}

For a thorough description of orbit analysis, 
we refer the reader to \cite{smile}. 

\subsection{Phase-space density estimation}

Several assumptions are made in the above approach with the most
important one being that the potential is frozen. It has been
suggested that gas cooling and star formation can change box orbits
into SAT \citep{Barnes1996, Hoffman2010}. It is not clear whether such
changes can be reversed. With a complementary approach, we compute a
numerical estimate of phase-space density $f(\mathbf{x}, \mathbf{v})$
using the {\sc EnBiD} code\footnote{The source code of {\sc EnBiD}
  available at \url{http://sourceforge.net/projects/enbid/}.}
\citep{Sharma2006} for DM particles. An SPH-like kernel smoothing is
applied to compute the phase-space density along an adaptive metric
with an anisotropic kernel. \cite{Vass2009} have shown that this
approach is able to reveal more detailed structures in DM halos in
phase-space, which are related to the subhaloes and newly formed tidal
streams from disrupted subhaloes, rather than using the spherically
averaged quantity $Q$($\equiv \rho/\sigma^3)$. As mixing (from phase
mixing, chaotic mixing, as well as violent relaxation) occurs, the
coarse-grained phase density $f$ decreases with time
\citep{Dehnen2005, Vass2009}.

%%%%%%%%%%%%%%%%% RESULTS %%%%%%%%%%%%%%%%%%sec:results

\begin{figure}
\begin{center}
\begin{tabular}{c}
\resizebox{0.5\columnwidth}{!}{\includegraphics{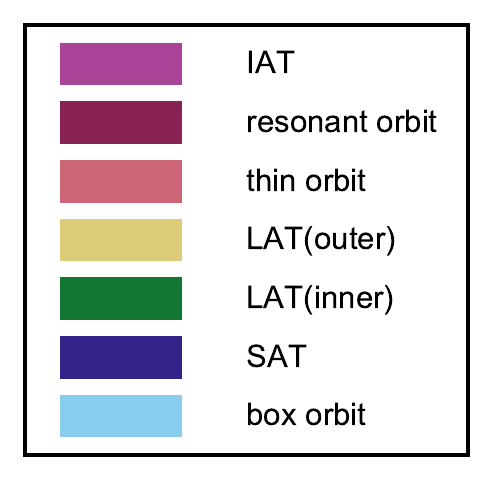}}\\
\vspace{-1cm}
\resizebox{0.85\columnwidth}{!}{\includegraphics{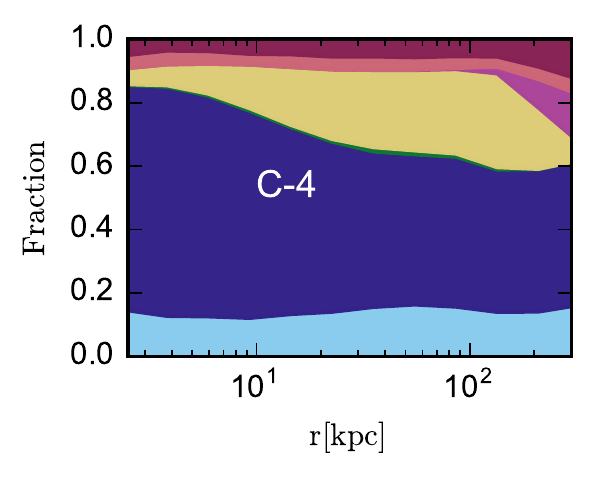}}\\
\resizebox{0.85\columnwidth}{!}{\includegraphics{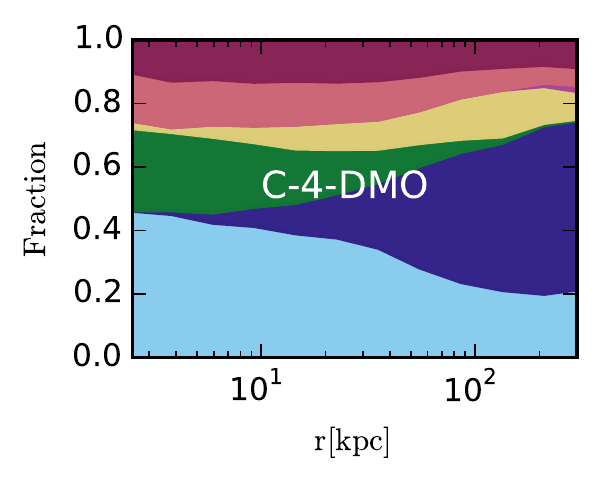}}\\
\end{tabular}
\end{center}
\caption{\label{fig:orbital_class} The orbital families and their radial distributions for the Aq-C-4 halo from both hydrodynamic ({\it middle panel}) and DMO ({\it bottom panel}) simulations, respectively. The color pie in the ({\it top panel}) represents the color scheme used to illustrate different orbital family. }
\end{figure}

\begin{figure}
\begin{center}
\resizebox{0.9\columnwidth}{!}{\includegraphics{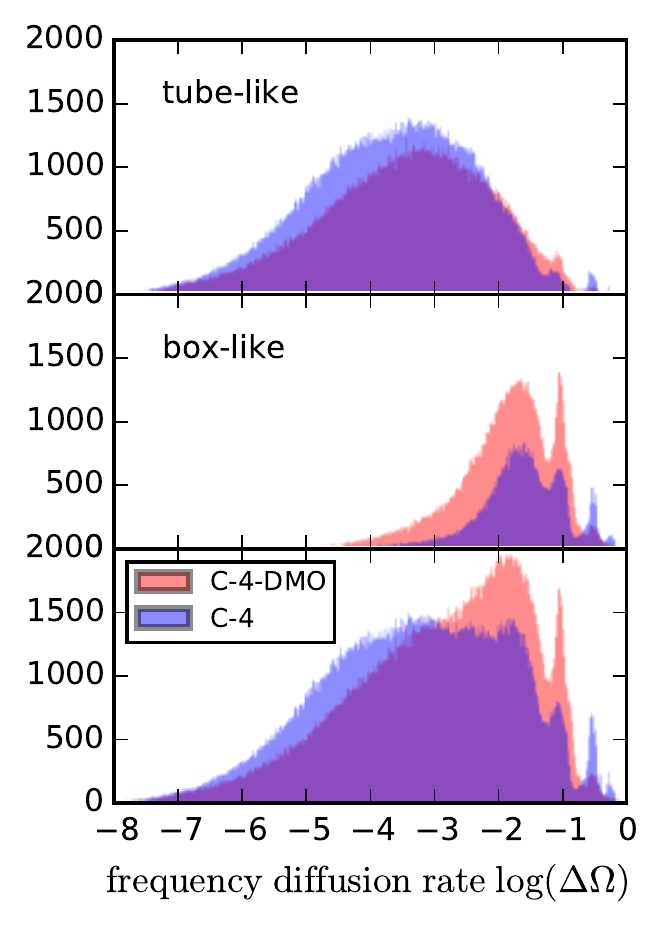}}
\end{center}
\caption{\label{fig:frequency_diffusion_rate} Histograms of frequency diffusion rates for tube-like orbits (SAT, LAT and resonant orbits, {\it top panel}), box-like orbits (thin and box orbits, {\it middle panel}) and the entire orbit family ({\it bottom panel}) of the Aq-C-4 halo from both hydrodynamic (in blue color) and DMO (in red color) simulations, respectively. There is a one percent  relative change between two halves of the interval for $\Delta\Omega \sim 0.01$.} Higher values of frequency diffusion rates correspond to more strongly chaotic orbits.
\end{figure}

\section{Results}
\label{sec:results}

\subsection{Dark matter halo shape}

In order to study the distribution of DM particles, we apply a
principal component analysis to compute the three axis parameters
${a}$, ${b}$ and ${c}$ based on the eigenvalues of the moment of
inertia tensor for all the DM particles within a given shell
\citep{Zemp2011}. The halo shape is quantified by the intermediate
to major axis ratio, ${b/a}$, and the minor to major axis ratio,
${c/a}$,  as well as a triaxiality parameter $T \equiv (a^2 -
b^2)/(a^2 - c^2)$.

A comparison of the DM halo shape of Aq-C between the hydrodynamic and 
DMO simulations is shown in Figure~\ref{fig:dmmaps}. The impact of hydrodynamics 
on the distribution of DM is clearly visible in the images of the projected density 
maps in the top panels.  For hydrodynamic simulations, haloes are more spherical
while the DMO counterparts show a clear boxy shape. This is confirmed
by the quantities ${b/a}$ and ${c/a}$ as a function of galactocentric
distance $r$. Within 100 kpc, ${b/a}$ remains above 0.8 for the
hydrodynamic simulation but around 0.6 for DMO simulations. The
triaxiality parameter $T$ clearly indicates that the inner regions of
the DM haloes in the hydrodynamic simulation are nearly spherical but
those of the DMO simulation comprise a prolate ellipsoid. It is also
evident that halo shape varies substantially as a function of $r$,
which indicates that the global shapes of DM haloes are not perfect
ellipsoids.

\subsection{Orbital properties: frequency map and orbital family}

As introduced in Section~\ref{subsec:orbitclassfication}, frequency maps can be used to analyse the
structure of the orbits based on the most prominent spectral lines in the frequencies of $\Omega_x$, $\Omega_y$ and
$\Omega_z$ \citep[][]{Papaphilippou1998, Valluri1998, Valluri2012}. We integrate individual orbits of DM particles using the mass
distribution from the simulation snapshots as in Figure~\ref{fig:dmmaps}, then apply Fourier transforms to the
orbit coordinates in each dimension.

The resulting frequency maps from our simulations are shown in Figure~\ref{fig:frequency_map}. 
Both SAT and LAT are densely populated in the hydrodynamic and DMO simulations. 
A large fraction of orbits in C-4-DMO are clustered around $\Omega_x/\Omega_y \sim 0.6$ and
$\Omega_x/\Omega_z \sim 0.8$, while the binding energy of each orbit
is around half of the maximum binding energy $\frac{1}{2}\max (E_{\rm
  binding})$.  Previous orbital classification has identified these
orbits as box orbits while our classification will further separate
thin orbits and resonant orbits from box orbits. For comparison,
the frequency map of the Aq-C-4 halo shows that this region is not
densely populated.

The bottom panels of Figure~\ref{fig:frequency_map} highlight
zoomed-in regions of the frequency space for the Aq-C-4 haloes. Many
orbits are thin orbits in the DMO simulation, as in the bottom right
panel, where many lines are heavily populated. In contrast, we find
far fewer orbits distributed along the same lines in the hydrodynamic
simulation. Moreover, we can also see that there are ``gaps'' along the
various resonant relations such that only segments of the lines are
populated in the DMO case. Physically, the ``gaps" correspond to
chaotic orbits where resonant relations overlap \citep[][]{BinneyTremaine, 
PriceWhelan2016}. While the orbits on the
resonant lines are stable, the other orbits are irregular orbits for which
the spectral method does not find well-defined frequencies.

With the frequencies computed as in Figure~\ref{fig:frequency_map}, we
can then assign orbital families to each individual orbit according to
the method described in Section~\ref{subsec:orbitclassfication}. 
Figure~\ref{fig:orbital_class} summarises the fraction of orbital 
families as a function of $r$ for the DM haloes from both 
hydrodynamic and DMO runs.

Figure~\ref{fig:orbital_class} shows the overall differences in the
final DM haloes in the orbits of DM particles. Most important, the
fraction of SAT orbits in the hydrodynamic simulation dominates the
other orbital families. Within the central 10 kpc, more than 50
percent of orbits are classified as SAT. The fraction of SAT orbits in
the C-4 simulation slowly decreases to 40 percent towards the outer
regions of the halo. The fraction of box orbits in the hydrodynamic
runs is subdominant, being slightly below 20 percent. For the DMO
simulation, the central regions are occupied by other orbital families
rather than SAT orbits. The relative weight of box orbits reaches up
to 40 percent within the central 10 kpc while the contribution from
inner LAT (which is almost absent in the hydrodynamic simulation),
thin and resonant orbits also increases substantially. For the latter
two orbital families, the combined contribution is around 20 percent
in the DMO simulation, which is in contrast to their relative light
weight in the hydrodynamic simulation. SAT orbits in the DMO
simulation start to appear only at galactocentric distances beyond
tens of kpc.

In the C-4 simulation, there is a very small fraction of orbits at
$\sim200$ kpc identified as IAT where $\Omega_x/\Omega_z \approx
1$. Such orbits are not stable in systems with perfect ellipsoidal
potentials. However in more realistic situations where the orientations
of the ellipsoidal shells are not well-aligned, such orbits can be
present.

Box orbits (and thin orbits) are the most important constituents of
triaxial systems \citep{Merritt1999}. The radial distributions of
orbital families in the two simulations are consistent with the halo
shapes shown in Figure~\ref{fig:dmmaps} as the triaxial halo in
the C-4-DMO simulation is dominated by box, resonant and thin orbits.

\subsection{Regular vs. chaotic orbits}

Regular orbits have definite frequencies in Fourier space. For
chaotic orbits, their frequencies are broader than distinct lines.  The
frequency diffusion rate \citep{Laskar1993} is a very useful tool for
studying the chaotic properties of each orbit. The definition of the
frequency diffusion rate in Equation~(\ref{eq:frequency_diffusion_rate})
quantifies the relative change of frequencies within the time interval
of hundreds of orbits. According to the analysis of \cite{smile},
the frequency diffusion rate is only an approximate indication of chaos as
there is an intrinsic scatter of $\sim0.5$ dex for
$\Delta\Omega$. Nevertheless, the frequency diffusion rate correlates
strongly with other chaos indicators such as the Lyapunov exponent
$\Lambda$. For $\Delta\Omega \sim 0.01$, there is a one percent 
relative change in the orbital frequencies between the two halves of the interval.
\footnote{We caution the reader that the large diffusion rate for many 
resonant/thin orbits can be caused by numerical artifacts. This happens when
two peaks have comparable magnitude, but one was slightly higher(lower) in the 
first half of orbit integration and lower(higher) in the second half. As a result, 
{\sc smile} will select two distinct leading frequencies in the two halves of 
the orbit integration leading to a large frequency diffusion rate according to 
Equation~(\ref{eq:frequency_diffusion_rate}).}

Figure~\ref{fig:frequency_diffusion_rate} shows the 
distribution of $\log(\Delta\Omega)$ for tube-like orbits such as SAT, 
LAT and resonant orbits,  box-like orbits (thin and box orbits), and all 
the orbits we integrated for the Aq-C-4 halo from both hydrodynamic and DMO simulations. 
Not surprisingly, most of the tube-like orbits are regular ones with
$\Delta\Omega$ well below 0.01. Chaotic orbits, those with
$\Delta\Omega$ above 0.01, are mostly associated with thin and box
orbits.  Interestingly, the fraction of chaotic orbits in the
hydrodynamic simulation is actually reduced rather than increased when
compared to the DMO simulation as shown in the lower panel for the
entire ensemble of orbits. We note there is a distinct peak of strong
chaotic orbits in the hydrodynamic simulation with $\log(\Delta\Omega)
\sim 0.5$.  We have verified that these orbits are associated with DM
particles at large galactocentric distances between 50 kpc and 250
kpc. Orbits in the central region of the DM halo in the hydrodynamic
run are mostly regular ones, which are SAT orbits characterised with
low-frequency diffusion rates.

\subsection{Angular momentum}

\begin{figure}
\begin{center}
\resizebox{\columnwidth}{!}{\includegraphics{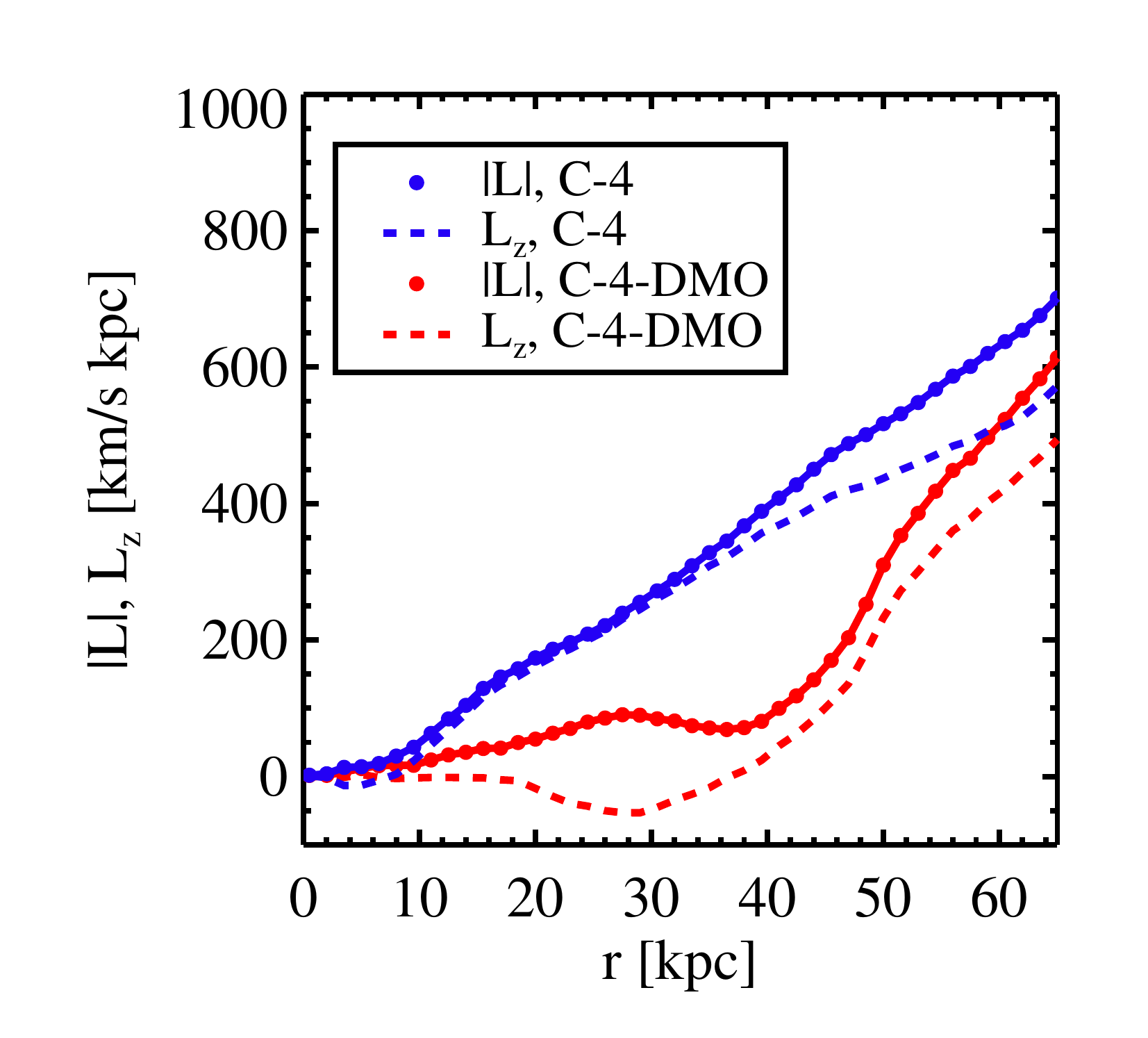}}
\end{center}
\caption{\label{fig:angular_momentum_transfer} A comparison of the mean angular momentum at different galactocentric distance $r$ of the Aq-C-4 halo from both hydrodynamic (in blue color) and DMO (in red color) simulations, respectively.  The mean total angular momentum, $|\mathbf{L}|$, is represented by the filled circles, while the mean values of angular momentum in the $z$ direction, $L_{z}$, is represented by the dashed curves. }
\end{figure}

In addition to halo shapes and orbital families, as we discussed in the previous sections, hydrodynamic physics has a global impact on the angular momentum distribution in DM haloes. In Figure~\ref{fig:angular_momentum_transfer}, we compare the mean angular momentum within concentric spherical shells of the Aq-C-4 halo from both hydrodynamic and DMO simulations. The angular momentum is calculated using the instantaneous position and velocity of each DM particle in the snapshot at $z = 0$. 

This plot shows a clear difference in both mean total angular momentum, $|\mathbf{L}|$, and the mean values of angular momentum in the $z$ direction, $L_{z}$ at all galactocentric distances between the hydrodynamic and DMO simulations. The central region of the DM halo in the hydrodynamic simulation shows clear net rotation as the mean values of $L_{z}$ are non-zero between $10$ and $30$ kpc, and it rises linearly up to $50$ kpc. This is also consistent with a large fraction of SAT orbits in the hydrodynamic run. For the DMO simulation, there is no coherent rotation along the $z$-axis within $20$ kpc as the $L_{z}$ curve lies close to zero. Beyond $20$ kpc, $L_{z}$ in DMO simulation switches from negative to positive at $\sim40$  kpc. 

Since the $z$-direction for the DM halo in the C-4 simulation is aligned with its stellar disc orientation, SAT orbits in the hydrodynamic simulation are either co-rotating or counterrotating with respect to the stellar disc. The sign of $L_z$ is positive beyond $10$ kpc, which means that most of the SAT orbits are corotating with the stellar disc. However within $10$ kpc, the fraction of counterrotating SAT orbits dominates over co-rotating SAT orbits, as the sign of $L_{z}$ is negative.

\begin{figure*}
\begin{center}
\resizebox{0.66\columnwidth}{!}{\includegraphics{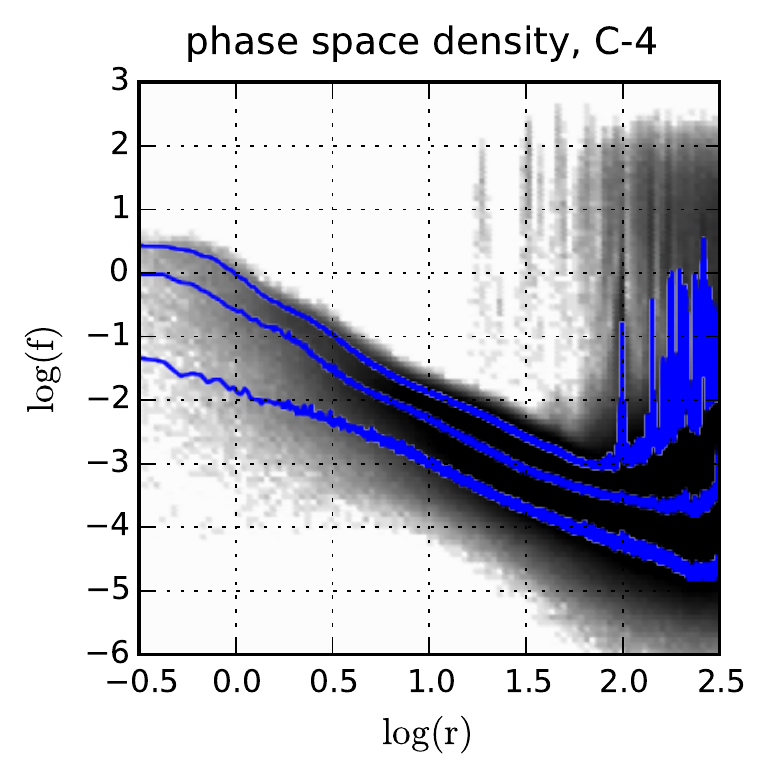}}
\resizebox{0.66\columnwidth}{!}{\includegraphics{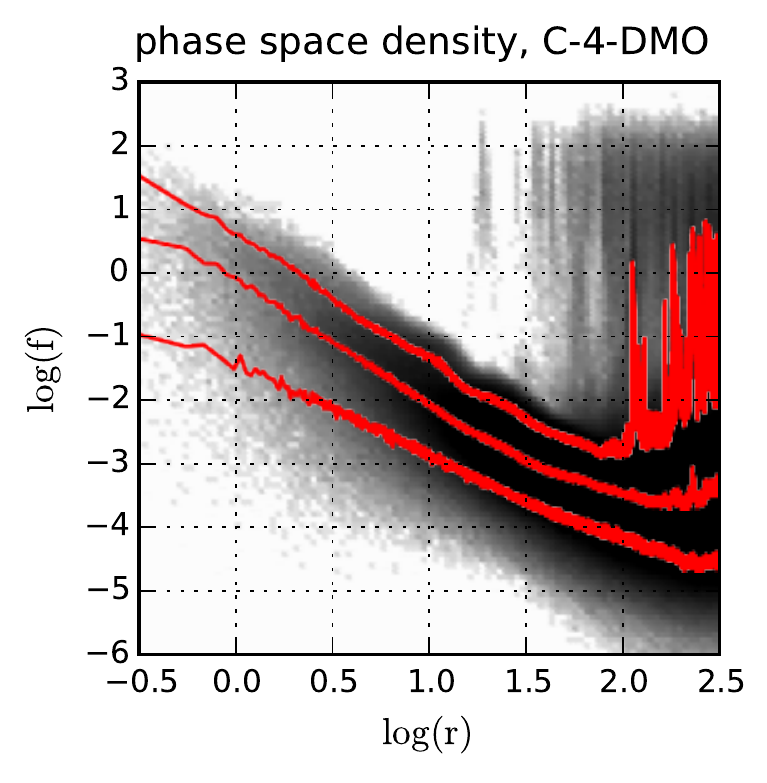}}
\resizebox{0.66\columnwidth}{!}{\includegraphics{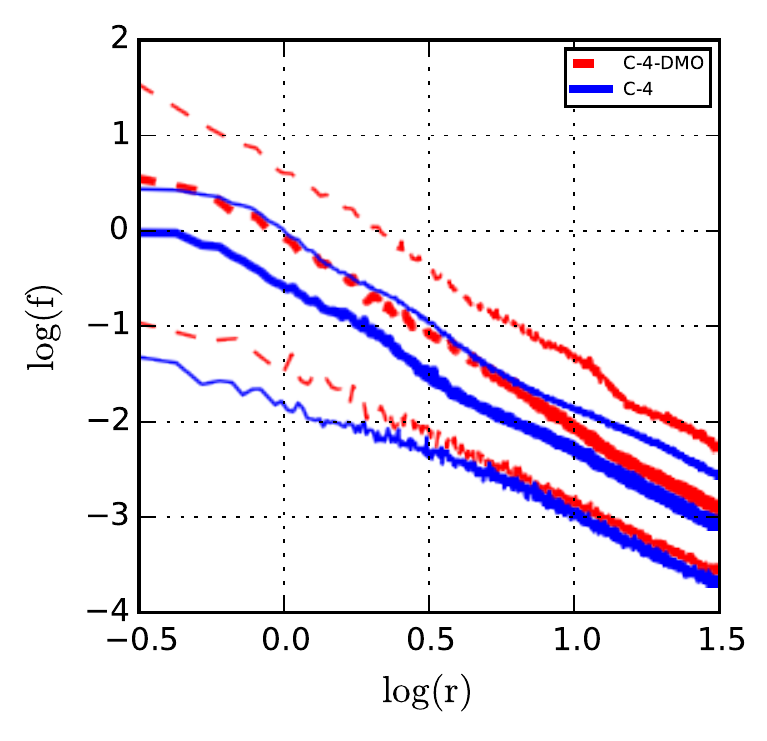}}
\end{center}
\caption{\label{fig:phase_space_density} A comparison of the coarse-grained phase-space density of DM particles at different galactocentric radius $r$ of the Aq-C-4 halo from both hydrodynamic and DMO simulations. The  phase-space density $f$ is given in code units [$10^{10} \rm{M_{\odot}}/h \ (Mpc/h)^{-3}\ (km/s)^{-3}$]. The solid colored lines in all panels represent the 10th, 50th and 90th percentiles. } 
\end{figure*}

\begin{figure*}
\begin{center}
\begin{tabular}{cc}
\resizebox{0.5\columnwidth}{!}{\includegraphics{orbit_class_color_scheme.pdf}}\\
\vspace{-1.0cm}
\resizebox{0.9\columnwidth}{!}{\includegraphics{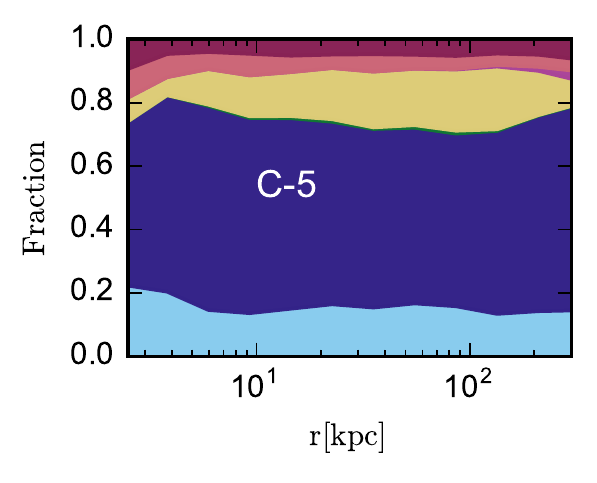}}
\resizebox{0.9\columnwidth}{!}{\includegraphics{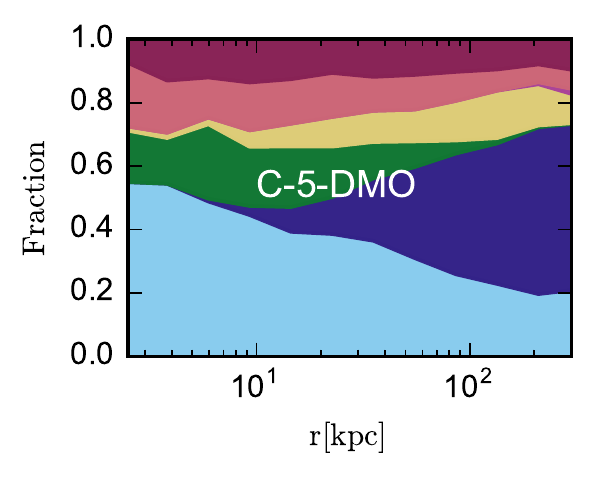}}\\
\resizebox{0.9\columnwidth}{!}{\includegraphics{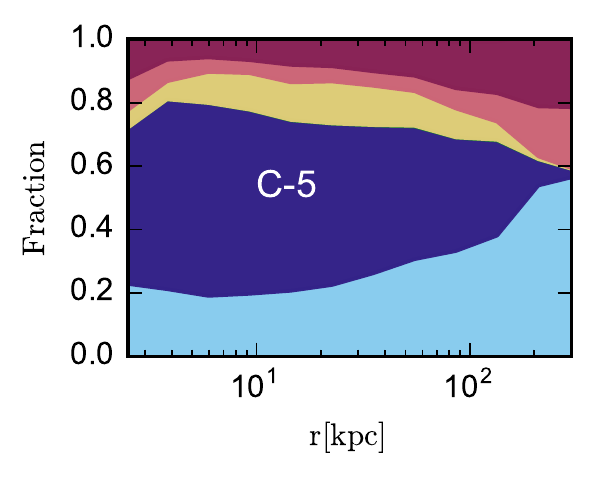}}
\resizebox{0.9\columnwidth}{!}{\includegraphics{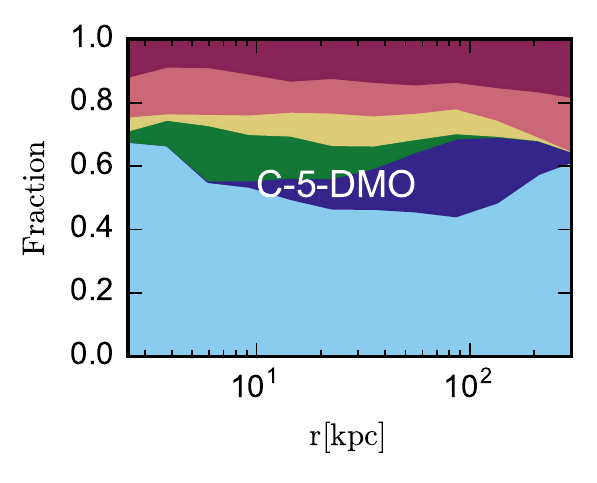}}\\
\end{tabular}
\end{center}
\caption{\label{fig:orbital_class_no_symmetry} The orbital families and their radial distributions for the Aq-C-5 halo from both hydrodynamic and DMO simulations, respectively.  The {\it top panels} show the orbital families with assumption of triaxial symmetry similar to the Aq-C-4 haloes in Figure~\ref{fig:orbital_class}, while the {\it bottom panels} without the triaxial symmetry assumption.}
\end{figure*}

\subsection{Phase-space density and the irreversible impact on DM due to baryons}

While orbital analysis provides a powerful diagnosis of the structure of
DM haloes, it is limited however due to the assumption that the
potential is ``frozen''.  In this section, we use a numerical estimate of
phase-space density as a complementary approach to study the evolution
of DM particles. The ``volume" that each particle occupies in phase
space contains important information about their state and subsequent
evolution.

Figure~\ref{fig:phase_space_density} shows the numerical estimate of the
coarse-grained phase-space density of DM particles as a function of
galactocentric distance $r$ for both hydrodynamic and DMO simulations. 

The overall radial distribution of the phase-space density of DM particles
in our simulations is close to that 
obtained by \cite{Vass2009}. High phase-space density regions
correspond to subhaloes and newly formed ``streams" from tidal
disruption, which are preferentially located at large galactocentric
distances. These are the ``spikes'' only found beyond
10 kpc. On the other hand, the central regions of DM haloes consist of
a well-mixed smooth component  \citep[see also][]{Diemand2008}.

In \cite{Zhu2016}, we compared the DM density profiles in the
hydrodynamic and DMO simulations for the central haloes. We found that
the difference between the two density profiles is fairly well captured by
the adiabatic contraction model of \cite{Gnedin2004}.  The analysis of
individual orbital properties in the present study indicates that the
impact of dissipational processes (from gas cooling, star formation
etc.) is more complicated. In particular, adiabatic changes of orbits
are reversible.

The numerical value of $f$ can be viewed as a combination of mass
density and local velocity dispersion. The right panel of
Figure~\ref{fig:phase_space_density} illustrates significant
differences between the hydrodynamic and DMO simulations. 
For DM particles in the central regions, their local velocity 
dispersion is greatly increased in the hydrodynamic simulation 
since the mass density in the hydrodynamic simulation 
is slightly higher than that in the DMO run \citep{Zhu2016}.
The median values of $f$ in the hydrodynamic 
simulation are lowered by 0.5 dex compared to that in the DMO halo.
In the very central regions, the 90th percentile $f$ of the
hydrodynamic halo is close to the median values of $f$ in the DMO
halo.

Such a difference in the phase-space density between the two
simulations suggests an acceleration of mixing processes leading to
a smooth component of DM haloes. \cite{Vass2009} have found that the
phase density of the central regions of DM haloes gradually decreases
from high redshift to low redshift. In addition, the difference in
phase-space densities also suggests that the impact is not
reversible. \cite{Debattista2008} and \cite{Valluri2010} show that if
the central baryon mass concentration slowly evaporates, DM haloes can
return to their original shape before the baryon mass is added. This
reversal cannot be achieved in our fully cosmological hydrodynamics
simulations if we slowly artificially remove the central mass
distribution.

One important reason that the impact on the DM distribution due to
baryons is irreversible owes to angular momentum exchange between the
baryon and DM components. In Figure~\ref{fig:angular_momentum_transfer}, 
the mean angular moment of each concentric DM shell in the hydrodynamic 
simulation is considerably larger than that of the DMO simulation. 
Moreover, within the central 30 kpc of the hydrodynamic DM halo, particles are
circulating coherently along the $z-$axis, which is also the
orientation of the stellar component.  Such a coherent rotation of DM
particles is not seen in the DMO simulation.

Another reason of the lower phase-space density $f$ in the hydrodynamic 
simulation lies in the combination of higher velocity dispersion and higher density of the 
DM particles compared with the DM run, and consequently the shorter dynamical 
time scale $t_{\rm dyn} \sim 1/\sqrt{G\rho}$. Unlike \cite{Debattista2008}, 
the scattering of box orbits due to the baryonic disc is very likely to be the root 
cause here because the fraction of the box orbits in the DMO halo is greatly reduced 
in the hydrodynamic simulation.

\cite{Valluri1998} suggested that many of the stellar orbits can
become chaotic in more axisymmetric systems when ``dissipation" is
added to triaxial potentials. This speculation is in the context of
understanding the systematic differences in the light distributions
between bright and faint elliptical galaxies. The same argument
applies to the shapes of DM haloes. Interestingly, the fraction of
chaotic orbits in the hydrodynamic simulation (shown in
Figure~\ref{fig:frequency_diffusion_rate}) is actually lowered rather
than increased relative to the dark matter only run. This is quite the
opposite of what \cite{Valluri1998} suggest. We note that the low
phase-space density in the C-4 simulation is consistent with an
acceleration of orbital dynamics in the discussion of
\cite{Valluri1998}. Somehow, a new order is established in the
hydrodynamic simulation.

%%%%%%%%%%%%%%%%%%%%%%%%%%%
%%%%%%%%%%  DISCUSSIONS      %%%%%%%%

\section{Discussion}
\label{sec:discussions}

In \cite{Zhu2016}, we emphasised the impact of baryonic processes on
the DM distribution, focusing on the evolution of subhaloes using a
high resolution cosmological hydrodynamic simulation of a Milky
Way-sized galaxy by \cite{Marinacci2014}. We showed that the combination
of re-ionization and stronger tidal disruption in the hydrodynamic
simulation leads to a different subhalo mass function and spatial
distribution. The low mass end of the subhalo function is reduced by
$\sim50\%$ in the hydrodynamic simulation due to those two
processes. The current study provides another rationale to use full
hydrodynamic simulations, which are more expensive than $N$-body runs,
to study the general properties of DM haloes.

\subsection{Numerical resolution and the assumption of triaxial symmetry}

We have run the same spectral analysis on a lower resolution
simulation in order to assess the impact of numerical
resolution. Figure~\ref{fig:orbital_class_no_symmetry} shows the
orbital classification of DM particles for C-5 haloes, with the left
panel indicating the hydrodynamic halo and the right panel the DMO
run. The fractions of different orbital families are quite consistent
between the C-5 and C-4 simulations even though the mass resolution of
the two suites differs by a factor of eight.  This suggests that our
results are numerically converged.

The mass distributions from the cosmological simulations are
represented with analytical functions in our orbital analysis. In this
study, we use spline functions in the radial direction and spherical
harmonics as the angular basis. While our choice of the number of
terms is able to match well the original potential which consists of
point masses, the choice can lead to some quantitative differences
nevertheless. The potential expansion we used also assumed triaxial
symmetry with only odd $l$ and $m$ terms. To assess the impact of this
assumption, we integrate the orbits in C-5 simulations without any
constraint on odd or even $l, m$ terms. Similar to the middle panel of
Figure~\ref{fig:orbital_class_no_symmetry}, the lower panel shows the
radial distribution of orbital families for the C-5 haloes but without
the assumption of triaxial symmetry.

In fact, the radial distributions of the fractions of each orbital
family are relatively insensitive to the assumption of triaxial
symmetry. In particular, within the central 30 kpc the fraction number of SAT
and box orbits in the hydrodynamic run is very close to that shown in
the middle panel. Most of the differences are seen in the outer halo,
where a significant of SAT orbits in the middle panel are replaced by
various resonant, thin and box orbits. Similar changes can also be
seen in the DMO halo where the SAT orbits in the triaxial symmetric
potential are replaced by box orbits. In addition, the unstable IAT
family (a negligible fraction for C-5 haloes in the middle panel of
Figure~\ref{fig:orbital_class_no_symmetry}) is absent when no triaxial
symmetry is assumed.

We also have examined the impact of number of radial grids, 
using $n_{\rm rad} = 20$ instead of our default value. The radial
distribution of each orbital family in this test largely remains the same
as the result obtained with $n_{\rm rad} = 12$. This is consistent with
\cite{smile} who found that the density and force approximations using 
splines with a modest $n_{\rm rad} (\in [10, 20])$ are sufficiently accurate.

\subsection{The assumption of ``frozen'' potential}

\begin{figure}
\begin{center}
\resizebox{0.9\columnwidth}{!}{\includegraphics{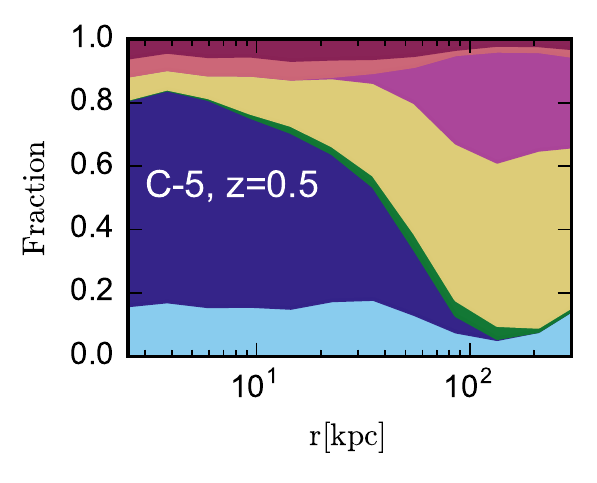}}\\
\vspace{-1.5cm}
\resizebox{0.9\columnwidth}{!}{\includegraphics{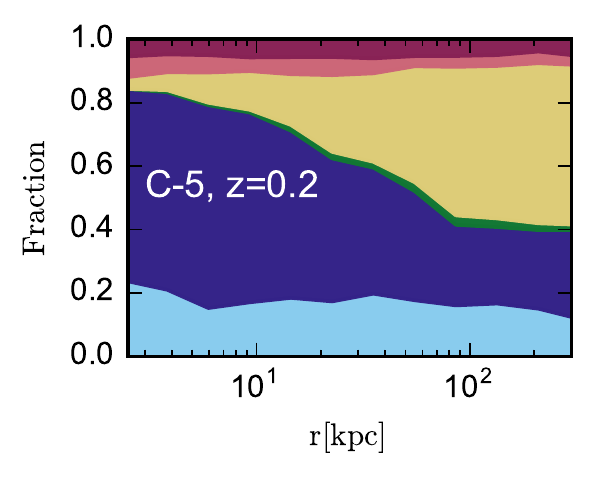}}\\
\vspace{-1.5cm}
\resizebox{0.9\columnwidth}{!}{\includegraphics{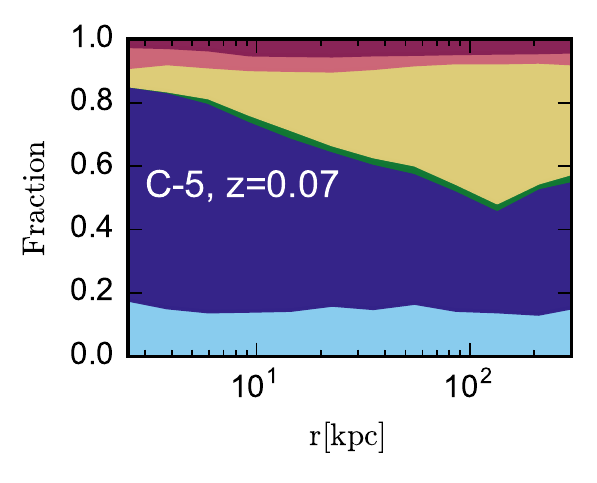}}\\
\caption{\label{fig:orbital_fraction_evolution} 
The radial distributions of the orbital families of the DM particles for the Aq-C-5 halo
at redshifts $z = 0.5$, $0.2$,  and $0.07$. There are substantial changes in the orbital 
family distributions, and hence in the DM halo shape, from $z = 0.5$ to $z = 0.07$, 
and then to $z = 0$ (upper left panel in Fig~\ref{fig:orbital_class_no_symmetry}). 
In particular, a large fraction of IAT orbits at $z = 0.5$ has disappeared, and the fraction 
of SAT orbits has increased significantly at $r\sim 100$ kpc. In contrast, 
the orbital family distribution in the inner halo shows much less evolution 
than that in the outer halo. }
\end{center}
\end{figure}

Another important assumption in our orbital classification is that the potential is frozen 
into a static one while in reality live DM halos are dynamically active because the mass assembly 
is ongoing process particularly in the outer halo \citep[e.g.,][]{Wang2011, More2015}. 

Figure~\ref{fig:orbital_fraction_evolution} shows the orbital families and their radial distributions 
for the Aq-C-5 halo in the hydrodynamic simulation at three separate redshifts $z = 0.5$, $0.2$, 
and $0.07$. For each snapshot, we repeat the same procedure of orbit integration and classification
with {\sc smile}. There are substantial changes in the orbital family distributions, and hence in the DM
halo shape, from $z = 0.5$ to $z = 0.07$, and then to $z = 0$. In particular, a large
fraction of IAT orbits at $z = 0.5$ has disappeared, and the fraction of SAT orbits has 
increased significantly in the outer halo since $z = 0.5$. 

Therefore, we conclude the assumption of a static potential is likely to have an impact on 
the orbital properties in the outer halo. One method to model of the evolution of the
potential is to use some time-varying set of coefficients, as adopted by \cite{Lowing2011}. 

%%%%% CONCLUSIONS %%%%%%
%%%%%%%%%%%%%%%%%%%

\section{Conclusions}
\label{sec:conclusions}

Current hydrodynamic simulations with comprehensive treatments of gas
cooling and star formation have shown substantial modifications to the
distribution of dark matter in galaxies. Following our study in
\cite{Zhu2016}, we have investigated the orbital properties of Milky
Way-sized DM haloes by integrating the orbits of a large number of DM
particles using spectral methods. Our key results are summarised as
follows:

\begin{itemize}
  \item DM haloes in hydrodynamic simulations appear more spherical than that of 
  DMO simulations. The changes of halo shape in hydrodynamic simulations are 
  reflected in the
  changes of orbital properties of individual particles. Box and thin orbits, which form 
  the backbone of triaxiality in DM haloes, are replaced by SAT orbits in the hydrodynamic case. 
  \vspace{0.5cm}

  \item
  While we find a substantial fraction of inner LAT orbits in the inner part of the DMO halo, 
  they are almost absent from the hydrodynamic halo.
  \vspace{0.5cm}  
  
  \item The fraction of chaotic orbits in the hydrodynamic halo is reduced compared to the DMO run. 
  This is a result of a large number of box and thin orbits in the DMO halo being replaced by 
  tube-like orbits, which are mostly regular.
  \vspace{0.5cm}

  \item There is a net, coherent rotation of DM particles in the central regions in the hydrodynamic DM halo,
which is absent in the DMO run. Within 10 kpc, the majority of DM particles on SAT obits are counter-rotating 
with respect to the stellar disc. 
   \vspace{0.5cm}

  \item In the central regions of the DM halo, the DM phase-space density is lowered by 0.5 dex when 
  hydrodynamic processes are included. Considering the changes of angular momentum and the changes
  of the orbital families, this strongly suggests that the impact of baryons on the DM distribution is not 
  reversible.
  
\end{itemize}

The spectral method and the classification of orbits using high
resolution cosmological simulations as in this study can be easily
extended to the core/cusp problem to offer additional insight into
recent investigations \citep{Madau2014, Onorbe2015, Chan2015,
Read2016}. We plan to carry out such an analysis on high resolution
cosmological simulations featuring bursty star formation using
explicit stellar feedback models.

\section*{Acknowledgements}
We thank the referee for a constructive and helpful report that improved the paper.
We thank Eugene Vasiliev for his {\sc smile} code and his 
comments and suggestions to the original manuscript.
We are grateful to Sanjib Sharma for making his codes {\rm EnBiD} publicly available. 
YL acknowledges support from NSF grants AST-0965694, AST-1009867,
and AST-1412719. We acknowledge the Institute For CyberScience at The Pennsylvania 
State University for providing computational resources and ser- vices that have contributed 
to the research results reported in this paper. The Institute for Gravitation and the Cosmos
is supported by the Eberly College of Science and the Office of the Senior Vice President for 
Research at the Pennsylvania State University. VS acknowledges support by the DFG 
Research Centre SFB-881 `The Milky Way System' through project A1, by the European 
Research Council under ERC-StG grant EXAGAL-308037, and by the Klaus Tschira Foundation.				

%%%%%%%%%%%%%%%%%%%%%%%%%%%%%%%%%%%%%%%%%%%%%%%%%%s
%%%%%%%%%%%%%%%%%%%% REFERENCES %%%%%%%%%%%%%%%%%%
% The best way to enter references is to use BibTeX:
\bibliographystyle{mnras}

%\bibliography{ref} 

%%%%%%%%%%%%%%%%%%%%%%%%%%%%%%%%%%%%%%%%%%%%%%%%%%
%%%%%%%%%%%%%%%%% APPENDICES %%%%%%%%%%%%%%%%%%%%%
%\appendix
%\section{Some extra material}
%If you want to present additional material which would interrupt the flow of the main paper,
%it can be placed in an Appendix which appears after the list of references.
%%%%%%%%%%%%%%%%%%%%%%%%%%%%%%%%%%%%%%%%%%%%%%%%%%

% Don't change these lines
\bsp	% typesetting comment
\label{lastpage}
\end{document}